\documentclass[a4paper]{aipproc}

\layoutstyle{6x9}

\begin{document}

\title [Two-Quark Correlations in the \\ Hard Electromagnetic Nucleon
Form Factors] {Two-Quark Correlations in the \\ Hard Electromagnetic
Nucleon Form Factors \footnote{Talk given by W. Schweiger at \lq\lq
Mesons and Light Nuclei `01\rq\rq, Prague, Czech Republic, July 2001}}


\author{M. Schw\"arz}{
  address={Institute of Theoretical Physics, University of 
  Graz, Universit\"atsplatz 5, \\A-8010 Graz, Austria},
  email={marc.schwaerz@uni-graz.at},
  thanks={ }
}

\iftrue
\author{W. Schweiger}{
  address={Institute of Theoretical Physics, University of 
  Graz, Universit\"atsplatz 5, \\A-8010 Graz, Austria},
  email={wolfgang.schweiger@uni-graz.at},
}

\fi

\begin{abstract}
The, so called, \lq\lq hard-scattering approach\rq\rq\ represents a
suitable framework for the perturbative treatment of exclusive
hadronic processes at large energies and (transverse) momentum
transfers.  In this context, diquarks can serve as a useful
phenomenological concept to model non-perturbative effects which are
still observable in the kinematic range accessible by present-day
experiments.  We outline how a description of baryons as quark-diquark
systems can be understood as an effective theory in the sense that the
pure quark hard-scattering approach is recovered in the limit of
asymptotically large momentum transfers.  Our arguments are based on a
reformulation of the hard-scattering formalism in terms of
quark-diquark degrees of freedom.  This reformulation provides the
exact form of photon- and gluon-diquark vertices and corresponding
vertex functions (diquark form factors) in the limit of asymptotically
large momentum transfers -- and thus also asymptotic constraints which
should be fulfilled by phenomenological quark-diquark models for hard
scattering.  As an application of this reformulation we present an
analysis of the hard electromagnetic nucleon form factors with respect
to their quark-diquark content.

\end{abstract}

\date{\today}

\maketitle

%
It is generally accepted that the, so called, \lq\lq hard scattering 
approach\rq\rq\ (HSA) gives the correct description of exclusive 
hadronic processes in the limit of asymptotically large (transverse) 
momentum transfers $Q^2$ (for an overview on the HSA see, e.g., 
Ref.~\cite{BL89}). The HSA is based on the factorization of hadronic 
amplitudes in process dependent, perturbatively calculable 
hard-scattering amplitudes and process independent distribution 
amplitudes (DAs) which contain the bound-state dynamics of the 
hadronic constituents. The corresponding analytical expression for 
one of the simplest exclusive quantities, the nucleon magnetic form 
factor, is given by 
\begin{eqnarray}\label{GM}
G_{M}^{N} (Q^2) & = & \int_0^1 \left[ \prod_{i=1}^3 dx_i \, 
\delta\left( 1 - \sum_{k=1}^3 x_k \right)\right] \left[ \prod_{j=1}^3 
dy_j \, \delta\left( 1 - \sum_{l=1}^3 y_l \right)\right] \nonumber \\ 
& & \times 
{\phi^N}^\dag(y_1, y_2, y_3; \tilde{Q}^2) T_H(x_1, \dots, y_3 ; Q^2)
{\phi^N}(x_1, x_2, x_3; \tilde{Q}^2) \; . 
\end{eqnarray}
In leading-order perturbation theory the hard-scattering amplitude 
$T_H$ is calculated on tree level for massless, collinear valence 
quarks. Two important consequences of this approximation are 
(fixed-angle) {\em scaling laws} and {\em hadronic-helicity 
conservation}. The latter means, in particular, that no quantitative 
statement can be made on the Pauli form factors $F_2^{N}$ within 
the (conventional) HSA. The distribution amplitude $\phi^N$ is a 
probability amplitude for finding the valence Fock state in the 
nucleon with the quarks carrying the fractions $x_i$ (or $y_i$) of 
the nucleon momentum and being collinear within an uncertainty 
$\tilde{Q}^2$ (which corresponds to the factorization scale). Its 
dependence on $\tilde{Q}^2$ is given by an evolution equation. The 
general solution of this evolution equation is known, the integration 
constants, however, have to be determined by non-perturbative means. 
In the asymptotic limit $\tilde{Q}^2 \rightarrow \infty$ the nucleon 
DA becomes particularly simple
\begin{equation}\label{DAas}
\phi_{\rm as}(x_1,x_2,x_3) \propto x_1 x_2 x_3 \; .
\end{equation}
Unfortunately it turns out that the leading-order perturbative results
for the nucleon magnetic form factors obtained with $\phi_{\rm as}$
are far away from the experimental data, even at the largest
experimentally accessible values of $Q^2$.  Reasonable results can be
obtained with asymmetric DAs, like the one proposed by Chernyak et
al.~\cite{COZ89}
\begin{equation}\label{DACOZ}
\phi_{\rm COZ}(x_1,x_2,x_3) \propto \phi_{\rm as}(x_1,x_2,x_3) (23.814 
x_1^2 + 12.978 x_2^2 + 6.174 x_3^2 + 5.88 x_3 - 7.098)\; .
\end{equation}
This DA fulfills QCD sum-rule constraints on its moments (for 
$\tilde{Q}^2 \approx 1$~GeV$^2$). It has, however, been objected that 
such an asymmetric DA leads to a situation which is problematic for a 
perturbative calculation~\cite{IL89}: If the momentum of a hadron is 
very unequally shared between the constituents the hard subprocesses 
are dominated by rather small gluon virtualities.

It is, nevertheless, not necessary to completely give up the HSA when 
working in the few-GeV momentum-transfer region. After having 
observed that strongly asymmetric DAs are required to reproduce 
experimental data it is, of course, tempting to associate this 
asymmetry with diquark clustering. This idea has been pursued in a 
series of papers~\cite{JaKroSchuSchw93, KroPiSchuSchw93, 
KroSchuGui96, KSPS97, BeSchw00}, in which a systematic study of 
various (photon-induced) exclusive reactions has been carried out 
within a HSA-based phenomenological quark-diquark model of baryons. 
The dynamics of the scalar and vector diquarks occurring within this 
model is determined by the gauge-boson diquark vertices and 
corresponding form factors. The form factors account for the 
composite nature of diquarks and have been chosen in such a way that 
the scaling behaviour of the pure quark HSA is recovered in the limit 
of asymptotically large momentum transfers. What one gains by 
modelling two-quark clusters by means of diquarks is that the gluons 
on the quark-diquark level are, on the average, harder than on the 
pure quark level.  The kinematic range in which a perturbative 
treatment is well justified is thus extended to smaller (overall) 
momentum transfers on the quark-diquark level.


A more formal justification of diquarks can be achieved by observing
that the diquark model -- and not only its scaling behaviour -- should
evolve into the pure quark HSA in the limit of asymptotically large
momentum transfers.  This suggests a reformulation of the pure quark
HSA in terms of quarks and diquarks.  Two obvious restrictions on this
reformulation are that the leading order hard-scattering amplitude on
quark-diquark level should only consist of tree graphs (like in the
pure quark HSA), and that the result of the reformulation should not
depend on whether quarks 1 and 2, 1 and 3, or 2 and 3 are grouped to a
diquark.  For the case of baryon form factors it has explicitly been
shown that such a reformulation is possible~\cite{Sch01}.  This can be
seen by splitting the calculation of the hard-scattering amplitude on
the pure quark level into two steps, namely first the calculation of
two-quark subgraphs with the two quarks being in a particular
spin-flavour state and afterwards the calculation of the full graphs
on quark-diquark level with the vertices and vertex functions obtained
for the two-quark subgraphs.  Correspondingly, the baryon DA has to be
decomposed into terms which belong to certain spin-flavour states of
the two-quark system $(i,j)$ consisting of quarks $i$ and $j$.  The
key observation is then that the groupings $(1,2)$, $(1,3)$, and
$(2,3)$ are connected by an appropriate interchange of the momentum
fractions which means that the integration variables in the
convolution integral, Eq.~(\ref{GM}), are only renamed.  Thus the
convolution integral, Eq.~(\ref{GM}), is independent on whether
quarks 1 and 2, 1 and 3, or 2 and 3 are grouped to a diquark (of
certain spin and flavour).  By summing furthermore all the tree graphs
on quark-diquark level over the groupings $(1,2)$, $(1,3)$, and
$(2,3)$ every graph is counted twice on quark level.  It thus suffices
to calculate all the tree graphs on quark-diquark level for a
particular (but arbitrary) grouping $(i,j)$ and to multiply the result
with a statistical factor $(3/2)$ to embrace all the tree graphs on
quark level.

\begin{table}[t!]
\setlength\AIPhlinesep{0pt}
\begin{tabular}{|c||r|r|r||r|r|r|}\hline & 
\multicolumn{3}{c||}{Asymptotic DA} &
\multicolumn{3}{|c|}{COZ DA} 
\rule[-4pt]{0pt}{15pt}\\ 
\hline Transition & $C_{\rm as}^{-1}Q^4 G_{\rm M}^{3, {\rm p}}$ &
$C_{\rm as}^{-1}Q^4 G_{\rm M}^{4, {\rm p}}$&{\rm Sum} & $C_{\rm
COZ}^{-1}Q^4 G_{\rm M}^{3, {\rm p}}$ & $C_{\rm COZ}^{-1}Q^4 G_{\rm
M}^{4, {\rm p}}$&{\rm Sum}
\rule[-6pt]{0pt}{19pt}
\\
\hline
$V_0^1[ud]\rightarrow V_0^1[ud]$ & 0.111 & 0.000 & 0.111
 & 0.016 & 0.003 & 0.019
\rule[-4pt]{0pt}{15pt} \\ \hline
$V_0^1[ud]\leftrightarrow V_0^0[ud]$ & 0 & 0 & 0
 & 0 & 0.004 & 0.004
\rule[-4pt]{0pt}{15pt} \\ \hline
$V_0^1[ud]\leftrightarrow S^1[ud]$ & 0 & 0 & 0
 & 0 & 0 & 0
\rule[-4pt]{0pt}{15pt}\\ \hline
$V_0^1[ud]\leftrightarrow S^0[ud]$ & 0 & -0.500 & -0.500
 & 0 & -0.063 & -0.063
\rule[-4pt]{0pt}{15pt}\\ \hline
$V_0^0[ud]\rightarrow V_0^0[ud]$ & 0 & 0 & 0
 & 0.001 & 0 & 0.001
\rule[-4pt]{0pt}{15pt}\\ \hline
$V_0^0[ud]\leftrightarrow S^1[ud]$ & 0 & 0 & 0
 & 0 & -0.002 & -0.002
\rule[-4pt]{0pt}{15pt}\\ \hline
$V_0^0[ud]\leftrightarrow S^0[ud]$ & 0 & 0 & 0
 & 0 & -0.011 & -0.011
\rule[-4pt]{0pt}{15pt}\\ \hline
$S^1[ud]\rightarrow S^1[ud]$ & 0 & 0 & 0
 & $\approx 0 $& $\approx 0$ & 0.001
\rule[-4pt]{0pt}{15pt}\\ \hline
$S^1[ud]\leftrightarrow S^0[ud]$ & 0 & 0 & 0
 & 0 & -0.002 & -0.002
\rule[-4pt]{0pt}{15pt}\\ \hline
$S^0[ud]\rightarrow S^0[ud]$ & 1.000 & 0 & 1.000
 & 1.000 & -0.005 & 0.995
\rule[-4pt]{0pt}{15pt}\\ \hline
$V^1_1[ud]\rightarrow V^1_1[ud]$ & 0 & -0.056 & -0.056
 & 0 & -0.002 & -0.002
\rule[-4pt]{0pt}{15pt}\\ \hline
$V^1_1[ud]\leftrightarrow V^0_1[ud]$ & 0 & 0 & 0
 & 0 & 0.036 & 0.036
\rule[-4pt]{0pt}{15pt}\\ \hline
$V^0_1[ud]\rightarrow V^0_1[ud]$ & 0 & 0 & 0
 & 0 & 0.007 & 0.007
\rule[-4pt]{0pt}{15pt}\\ \hline
$V_0^1[uu]\rightarrow V_0^1[uu]$ & -0.111 & 0 & -0.111
 & -0.016 & 0.027 & 0.012
\rule[-4pt]{0pt}{15pt}\\ \hline
$V^1_0[uu]\leftrightarrow S^1[uu]$ & 0 & 0 & 0
 & 0 & -0.003 &-0.003
\rule[-4pt]{0pt}{15pt}\\ \hline
$S^1[uu]\rightarrow S^1[uu]$ & 0 & 0 & 0
 & $\approx 0 $& 0.005 & 0.004
\rule[-4pt]{0pt}{15pt}\\ \hline
$V^1_1[uu]\rightarrow V^1_1[uu]$ & 0 & -0.444 & -0.444
 & 0 & -0.013 & -0.013
\rule[-4pt]{0pt}{15pt}\\ \hline & 
\multicolumn{2}{c|}{Total}& 0 &
\multicolumn{2}{|c|}{Total}& 0.985 
\rule[-4pt]{0pt}{15pt}\\ \hline
\end{tabular}
\caption{ Decomposition of the proton magnetic form factor into
diquark contributions for the asymptotic proton DA  
\protect{(Eq.~(\ref{DAas}))} and the proton DA proposed by Chernyak 
et al. \protect{(Eq.~(\ref{DACOZ}))}, respectively.  $S^{I}[q_1 q_2]$ 
and $V^{I}_{h}[q_1 q_2]$ denote scalar and vector 
diquarks (with isospin $I$ and helicity $h$), respectively, 
consisting of quarks $q_1$ and $q_2$.  The various diquark 
contributions are further
decomposed into a 3- and 4-point part, $G_{\rm M}^{3, {\rm p}}$ and
$G_{\rm M}^{4, {\rm p}}$, depending on whether one or two gauge bosons
couple to the diquark.  The constants $C_{\rm as}= 4.646 \times 
10^{-2}$ and $C_{\rm COZ}= 1.266$ are chosen in such a way that the 
largest contribution, i.e. the $S^0[ud]\rightarrow S^0[ud]$ 
transition, becomes~1. \newline 
\hspace{1.0cm}
\label{tab1}}
\end{table}

Without explicitly calculating the (sometimes rather complicated) 
vertex structure of the gauge-boson diquark vertices this 
reformulation can already be used to analyze which diquark states are 
(dis)favoured by a particular baryon distribution amplitude. The 
diquark contents of the proton magnetic form factor for the asymptotic 
DA, Eq.~(\ref{DAas}), and the COZ-DA, Eq.~(\ref{DACOZ}), are compared 
in Tab.~\ref{tab1}. In both cases the most important contribution 
comes from the $S^0[ud]$ diquark. In case of the asymptotic DA this 
contribution is, however, completely compensated by contributions 
from vector diquarks and also transitions between scalar and vector 
diquarks. The asymmetry of the COZ-DA, on the other hand, leads to a 
strong suppression of all the other contributions in favour of the 
$S^0[ud]$ diquark. A similar observation can also be made for the 
neutron magnetic form factor~\cite{Sch01}. This suggests that a 
nucleon at intermediate momentum transfers appears approximately like 
a quark-scalar-diquark system.

Up to this point the term \lq\lq diquark\rq\rq\ has just referred to a
spin-flavour state of two quarks in a baryon.  Diquarks start to play
a dynamical role if one assumes that two quarks in a particular
spin-flavour state become strongly correlated at moderately large
momentum transfers such that they cannot be resolved but appear rather
as a quasi-elementary particle.  This should be reflected by the
corresponding diquark form factors which acquire a value of ${\cal
O}(1)$ at small enough momentum transfers when the diquark appears
nearly pointlike.  If all the correlations between the valence quarks
of a baryon are absorbed into the diquark form factors -- which means
that all deviations from the asymptotic form of the baryon DA are
shifted to the quark-quark DA of the diquark -- the perturbative
calculation of the diquark form factors may become problematic, since
the exchanged gluon becomes relatively soft.  The way out is to
consider the diquark form factors (at intermediate momentum transfers)
as non-perturbative quantities which have to be parameterized
phenomenologically.  In order to finally arrive at a diquark model for
hard exclusive scattering which can be understood as an effective
model in the sense that it reproduces the results of the pure quark
HSA in the limit of asymptotically large momentum transfers one should
then take the gauge-boson diquark vertices resulting from the
reformulation of the pure quark HSA and use the results for the
corresponding vertex functions as asymptotic constraints for the
parameterization of the diquark form factors.  A corresponding program
is presently carried out.

\end{document}